\documentclass{elsart1p}
\usepackage{graphics}
\usepackage{epsfig}
\usepackage{graphicx}
\usepackage{amssymb}
\begin{document}
\begin{frontmatter}
%
%
%
\title{Low-energy QCD and hadronic structure}
%
%
\author{Wolfram Weise}\footnote{Work supported in part by BMBF, GSI and by the DFG cluster of excellence Origin and Structure of the Universe.}
\address{Physik-Department, Technische Universit\"at M\"unchen, D-85747 Garching, Germany}
\begin{abstract}
Recent developments and selected topics in low-energy QCD are summarized, from chiral effective field theory to systems with strange and charm quarks, from lattice QCD to precision experiments.
\end{abstract}
%
%
\end{frontmatter}
%
\section{\label{prelude}Prelude: scales and symmetry breaking patterns}
{\it 1.1 The quark mass hierarchy}.
QCD without quarks, or with infinitely heavy quarks, has a remarkable feature: it is a gauge field theory with no parameters. A single scale, $\Lambda_{QCD}$, is introduced solely through renormalization. 
The actual hierachy of quark masses \cite{pdg08} introduces additional scales, ranging from the lightest quarks with $m_d\simeq 3-7$ MeV and $m_u/m_d \sim 0.3-0.6$, via the strange quark with $m_s \simeq 70 - 120$ MeV (each taken at a renormalization scale $\mu \simeq 2$ GeV) and the heavier charm quark with $m_c \simeq 1.3$ GeV to the heaviest quarks ($m_b \simeq 4.2$ GeV, $m_t \simeq 174$ GeV). The low-energy, long wavelength limit of QCD is expressed in terms of different types of effective field theories, depending on the quark masses involved:

i) Low-energy QCD with light quarks is realized in the form of an effective field theory with spontaneously broken chiral symmetry, expanded around the (chiral) limit of massless quarks. The active degrees of freedom are the Nambu-Goldstone bosons of this spontaneously broken symmetry, identified with the lightest pseudoscalar mesons. The small expansion parameters are the quark masses, $m_q$, together with low energy/momentum as compared to the characteristic chiral symmetry breaking scale, $4\pi f_\pi \sim 1$ GeV (with the pion decay constant $f_\pi \simeq 0.09$ GeV).

ii) Low-energy QCD with heavy quarks is also realized in the form of an effective field theory, non-relativistic QCD, in which a systematic expansion is now controlled by the inverse quark mass, $1/m_Q$. This expansion is valid for the heaviest ($b$- and $t$-) quarks and, to a lesser extent, also for $c$-quarks. In this report we touch upon recent developments in the physics with charmed quarks but do not digress on systems with the heaviest quarks.

Over the years, both these types of effective field theories have been established as approriate frameworks, within their ranges of applicability, for hadron physics and hadronic interactions. In conjunction with lattice simulations \cite{aoki09} utilizing steadily increasing computing power, these are now reliable tools for dealing with the non-perturbative areas of QCD.  
    
{\it 1.2 Phases of QCD}. The two prominent phenomena characteristic of low-energy QCD, confinement and spontaneous chiral symmetry breaking, are governed by basic symmetry principles:

i) An exact symmetry associated with the center $Z(3)$ of the local $SU(3)$ color gauge group is realized in pure gauge QCD, i.e. for {\it infinitely heavy} quarks, the limit in which gluons are the only active degrees of freedom. 
The deconfinement transition in this limiting situation is a 1st order phase transition. In the high-temperature, deconfined phase of QCD the $Z(3)$ symmetry is spontaneously broken, with the Polyakov loop acting as order parameter.

ii) Chiral $SU(N_f)_R\times SU(N_f)_L$ symmetry is an exact global symmetry of QCD with $N_f$ {\it massless} quark flavors. 
In the low-temperature (hadronic) phase this symmetry is spontaneously broken down to the flavor group $SU(N_f)_V$ (the isospin group for $N_f = 2$ and the ``eightfold way" for $N_f = 3$). As consequence there exist $2N_f - 1$ pseudoscalar Nambu-Goldstone bosons and the QCD vacuum is non-trivial. It hosts  quark condensates $\langle \bar{q} q \rangle$ which act as chiral order parameters. In the limit of $N_f = 2$ massless quarks the transition from spontaneously broken chiral symmetry to its restoration in the Wigner-Weyl realization is a 2nd order phase transition signaled by 
the ``melting" of the quark condensate. For $N_f = 3$ massless quarks this phase transition is first order. 

Yet there is no a priori reason why these two distinct symmetry breaking scenarios should be fundamentally connected. Confinement of massless quarks is understood to imply spontaneous chiral symmetry breaking, whereas the reverse is not necessarily true. Whether and under which conditions the chiral and deconfinement transitions coincide is therefore a crucial question.

\begin{figure}[htb]
\begin{minipage}[t]{6cm}
\includegraphics[width=6cm]{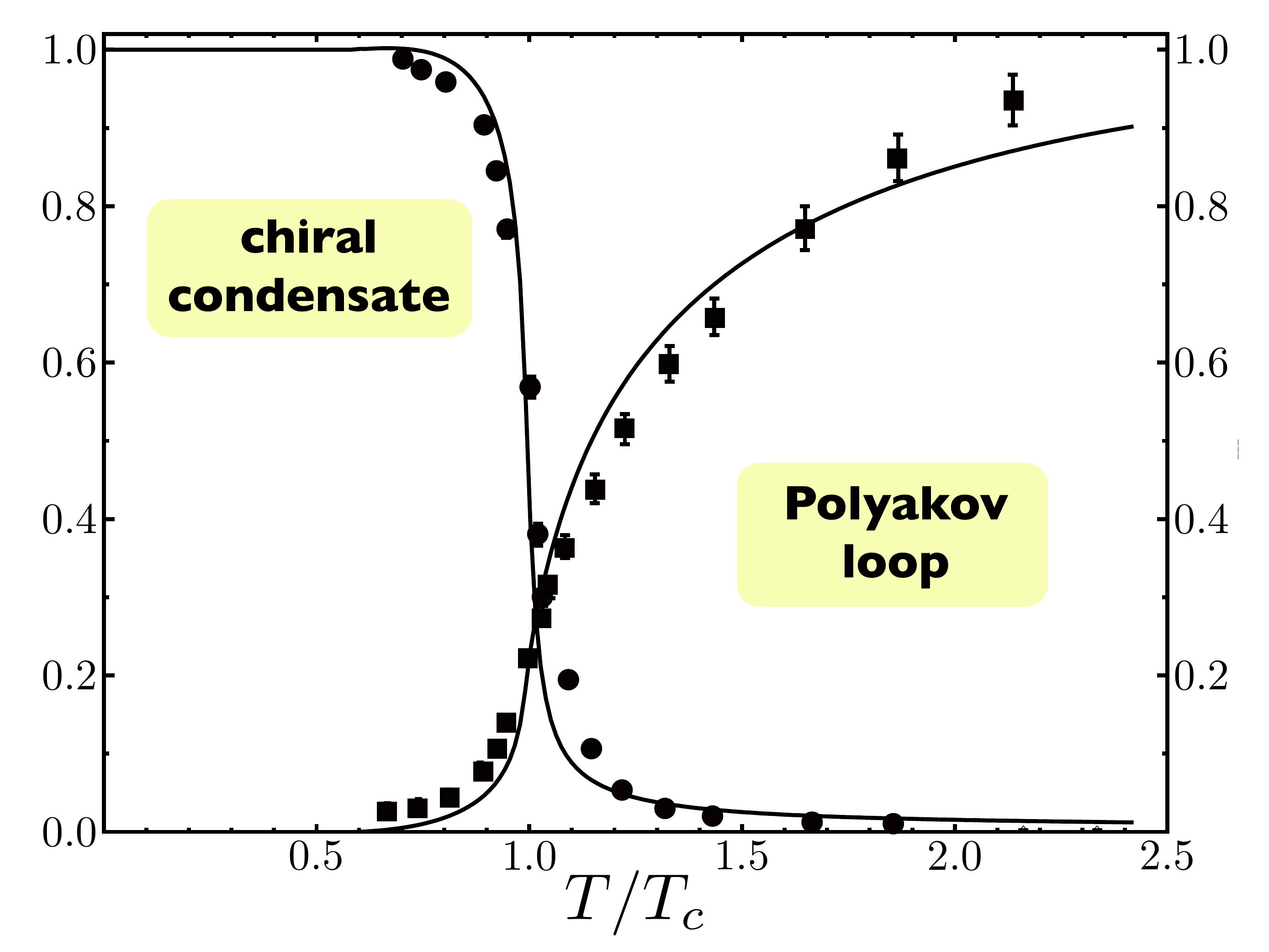}
\caption{Chiral condensate $\langle \bar{q}q\rangle$ (left) and Polyakov loop (right) as function of temperature in units of $T_c \simeq 190$ MeV. Lattice QCD results \cite{cheng08} are compared with a PNJL (Polyakov, Nambu $\&$ Jona-Lasinio) model calculation \cite{roessner07}. } 
\label{fig:1}
\end{minipage}
\hspace{\fill}
\begin{minipage}[t]{6cm}
\includegraphics[width=6cm]{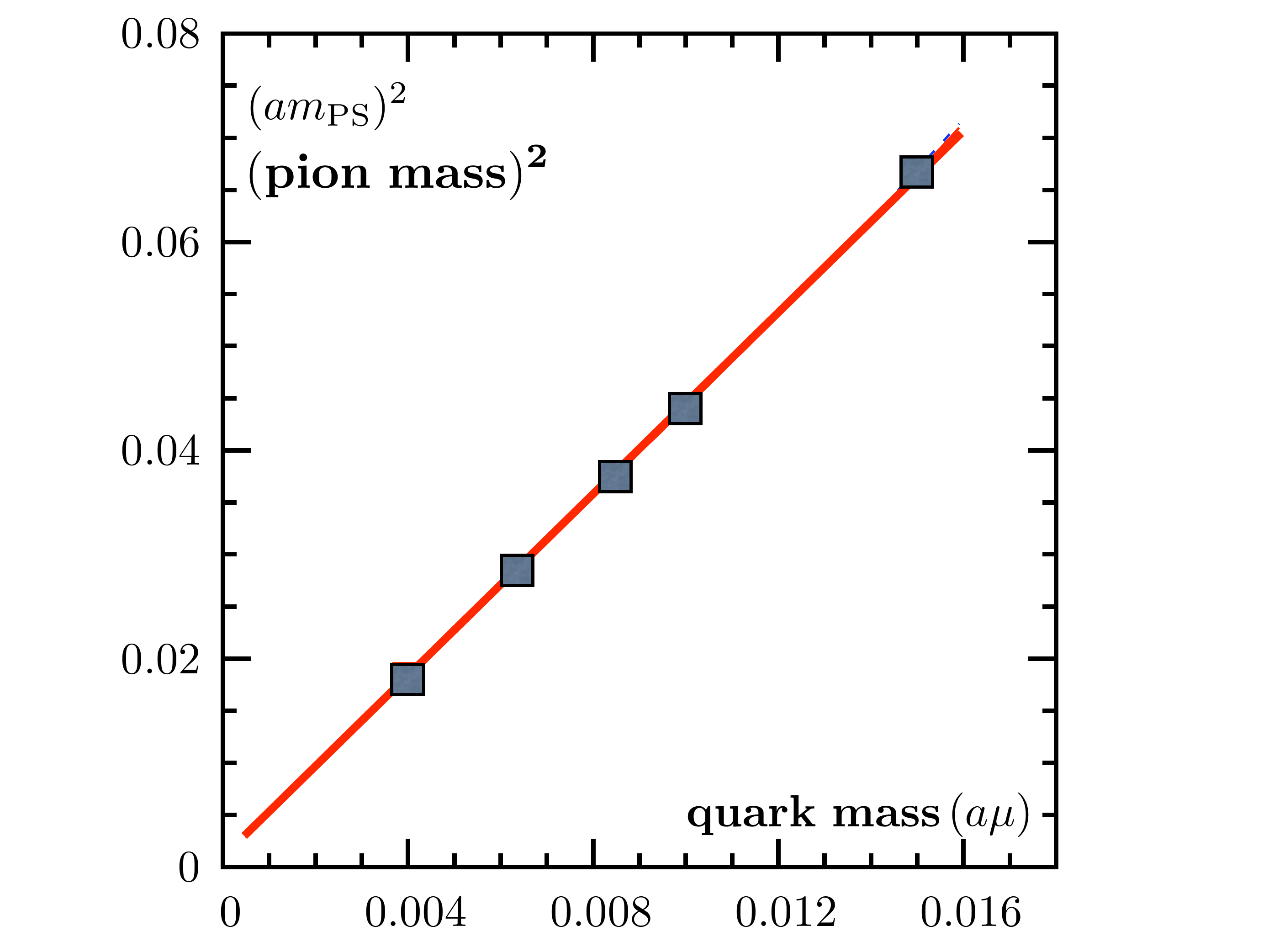}
\caption{Squared pion mass versus quark mass (both in lattice units) from lattice QCD (squares) \cite{boucaud07} and chiral perturbation theory (solid line). } 
\label{fig:2}
\end{minipage}
\hspace{\fill}
\end{figure}

Both deconfinement and chiral transitions become continuous crossovers when quarks are implemented and chiral symmetry is explicitly broken by non-zero quark masses. Recent results from $N_f = 2+1$ lattice QCD thermodynamics \cite{cheng08} do indicate that these two transitions have their steepest slopes at approximately the same critical temperature of about 190 MeV, as shown in Fig.\ref{fig:1}. This entanglement of chiral and deconfinement transitions is also seen in recent calculations which combine the Nambu $\&$ Jona-Lasinio model with Polyakov loop dynamics (the PNJL model) \cite{roessner07}. It is not observed, however, in earlier lattice computations \cite{aoki06} which suggest a splitting between chiral and deconfinement temperatures by about 20 MeV. It is important that this issue be clarified in the near future. 

\section{Low-energy QCD with light quarks}

{\it 2.1 The meson sector.} Lattice QCD and chiral effective field theory applied to the light-meson sector have now reached a remarkable degree of quantitative consistency. Together with high-precision experiments,  low-energy QCD has become a quantitative science.

A first and foremost  test of the spontaneous chiral symmetry breaking scnenario in QCD concerns fundamental properties of the pion: its mass $m_\pi$ and its decay constant $f_\pi$. At next-to-leading order (NLO)
in chiral perturbation theory \cite{gasser84}, 
\begin{equation}
m_\pi^2 = m^2\left[1+{m^2\over 32\pi^2 f^2}\ln{m^2\over \Lambda_3^2} + {\cal O}(m^4)\right],~~f_\pi= f\left[1-{m^2\over 16\pi^2 f^2}\ln{m^2\over \Lambda_4^2} + {\cal O}(m^4)\right]
\label{pionmass}
\end{equation}
with $m^2 f^2 = -m_q\langle\bar{q}q\rangle$ at leading order where $f$ is the pseudoscalar decay constant in the chiral limit,
$m_q \rightarrow 0$. A typical result from lattice QCD, shown in Fig.\ref{fig:2}, demonstrates the accuracy to which the linear relation $m_\pi^2 \sim m_q$ between squared pion mass and quark mass, characteristic of a Nambu-Goldstone boson, is fulfilled with small NLO corrections.  

\begin{figure}[htb]
\begin{minipage}[t]{6.5cm}
\includegraphics[width=6cm]{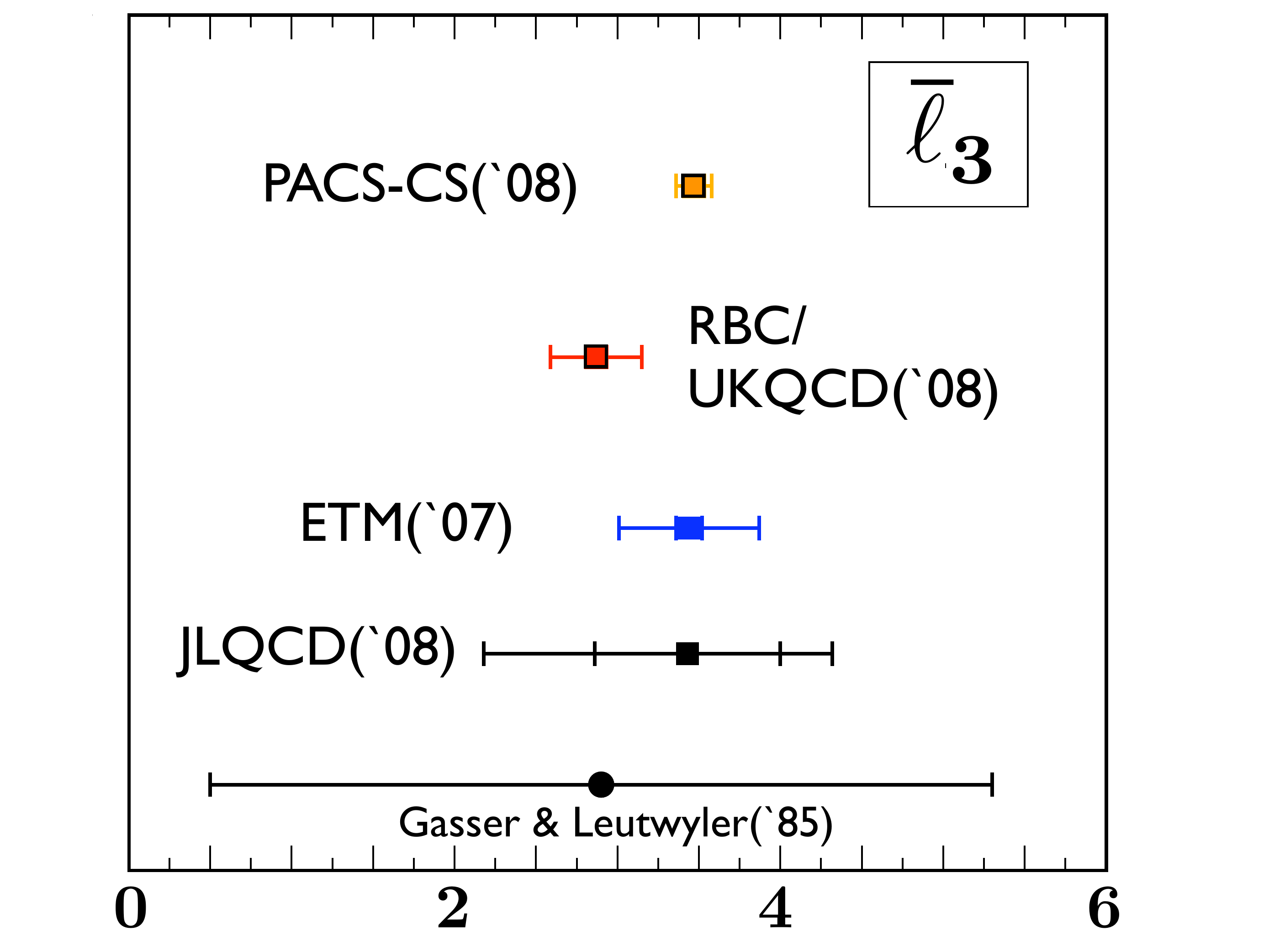}
\caption{Low-energy constant $\bar{\ell}_3$ from lattice QCD \cite{allton08} and a chiral theory \cite{gasser84}.}
\label{fig:3}
\end{minipage}
\hspace{\fill}
\begin{minipage}[t]{6.5cm}
\includegraphics[width=6cm]{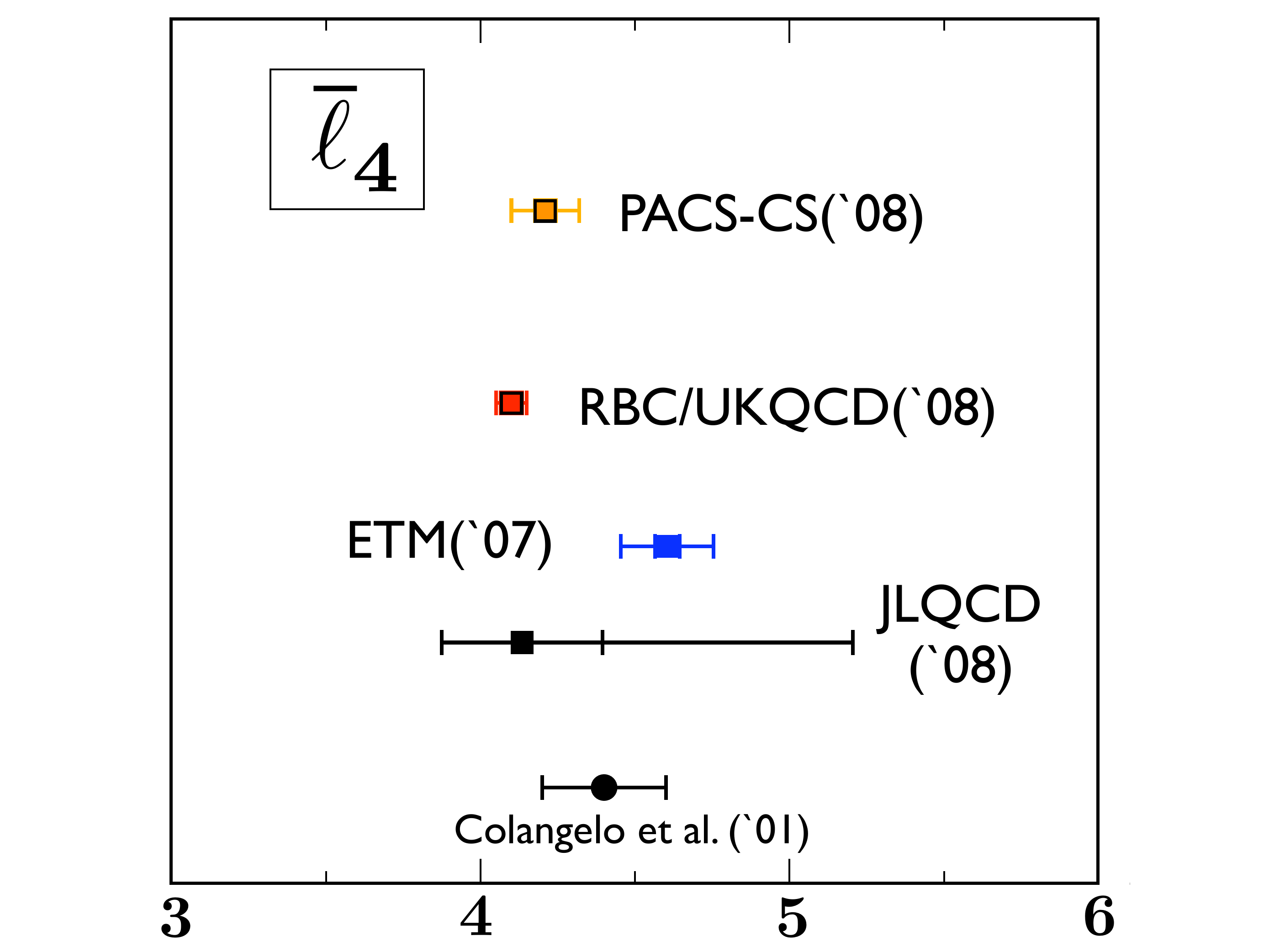}
\caption{Low-energy constant $\bar{\ell}_4$ from lattice QCD \cite{allton08} and chiral theory \cite{colangelo01}.}
\label{fig:4}
\end{minipage}
\end{figure}

The two low-energy constants appearing in  Eq.(\ref{pionmass}) are usually expressed in the form $\bar{\ell}_3 \equiv \ln(\Lambda_3^2/m_\pi^2)$ and  $\bar{\ell}_4 \equiv \ln(\Lambda_4^2/m_\pi^2)$. These quantitites, together with a series of other low-energy constants, are now determined with high precision from lattice computations (see e.g. \cite{allton08}). A survey of lattice results for   $\bar{\ell}_{3,4}$ is displayed in Figs.\ref{fig:3},\ref{fig:4}.  

The low-energy constants $\bar{\ell}_{3,4}$ enter sensitively in the $\pi\pi$ scattering amplitude close to threshold. A much improved analysis of the two-pion subsystem in the final state of the $K_{e4}$ decays $K^\pm \rightarrow \pi^+\pi^-e^\pm\nu$, performed by the NA48/2 collaboration at CERN \cite{na48/2}, has accurately determined the isospin $I=0,2$ $\pi\pi$ scattering lengths: $a_0 = (0.218\pm 0.013)$m$_\pi^{-1}$ and $a_2 = (-0.046\pm 0.009)$m$_\pi^{-1}$, in perfect agreement with the theoretical values $a_0 = (0.220\pm 0.005)$m$_\pi^{-1}$ and $a_2 = (-0.044\pm 0.001)$m$_\pi^{-1}$ \cite{colangelo01}.

\begin{figure}[htb]
\begin{minipage}[t]{6.5cm}
\includegraphics[width=6cm]{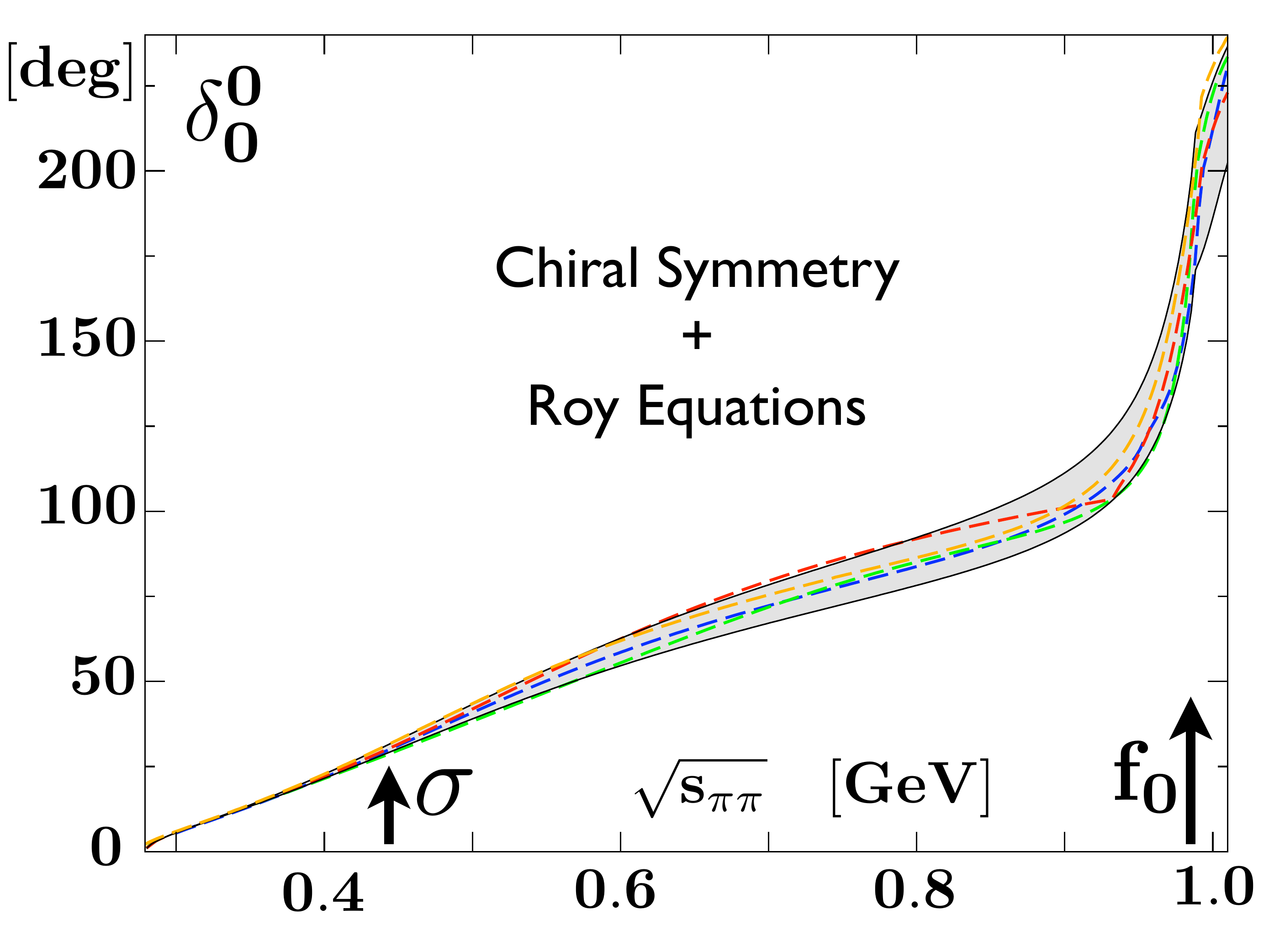}
\caption{Pion-pion isoscalar s-wave phase shift from chiral symmetry and Roy equations \cite{caprini06}.}
\label{fig:5}
\end{minipage}
\hspace{\fill}
\begin{minipage}[t]{6.5cm}
\includegraphics[width=6cm]{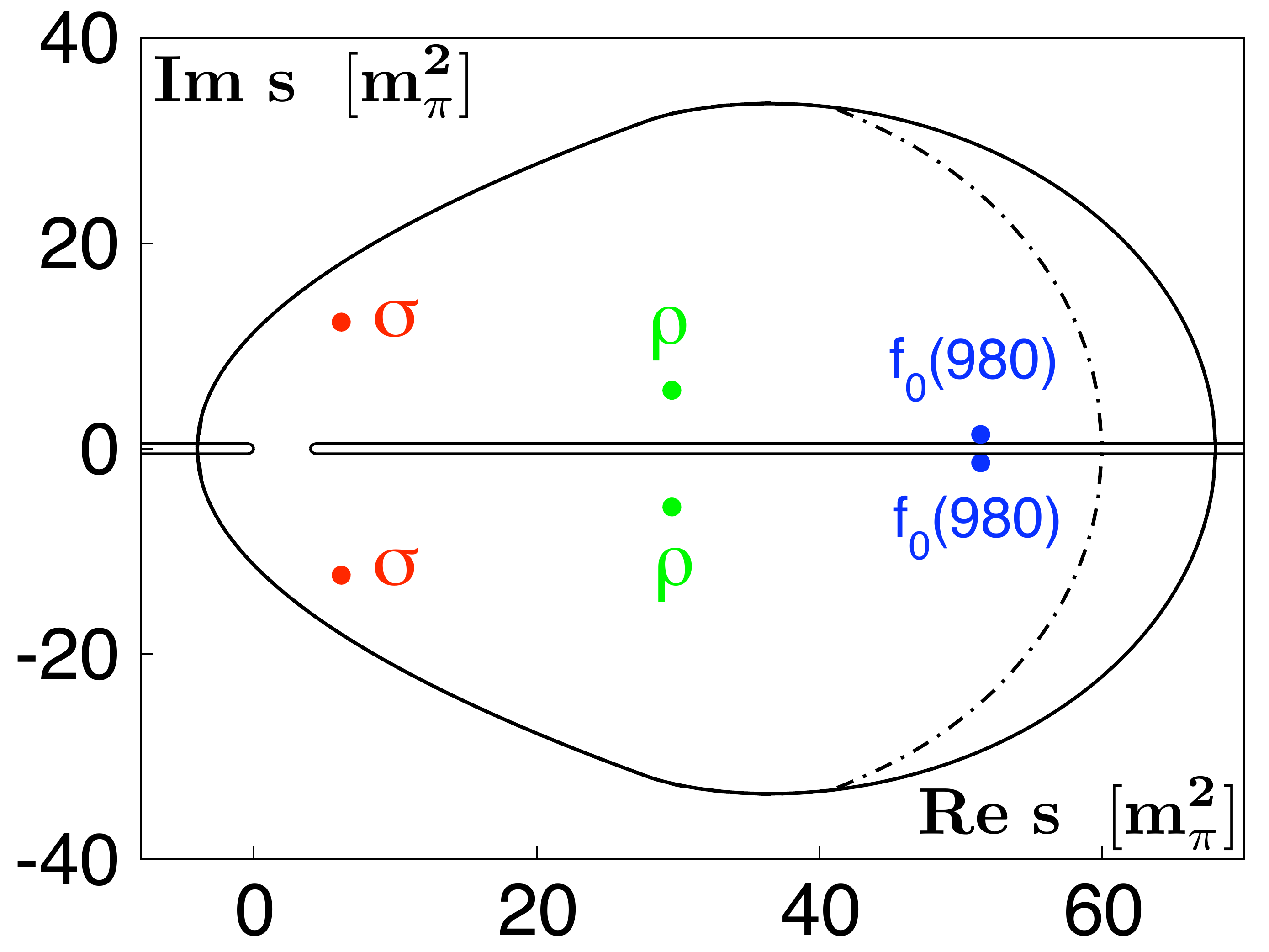}
\caption{Pattern of singularities in $\pi\pi$ scattering (from \cite{caprini06}).}
\label{fig:6}
\end{minipage}
\end{figure}

Another basic issue that has been around for decades and has recently been clarified is the scalar-isoscalar ``sigma" pole in pion-pion scattering. Detailed calculations of the $I=0$ s-wave $\pi\pi$ amplitude and phase shift (Fig.\ref{fig:5}) using chiral symmetry and Roy equations \cite{caprini06} have established a $\sigma$ pole with mass and width
\begin{equation}
M_\sigma = 441^{+16}_{-8}~{\rm MeV}~~,~~~~~~~\Gamma_\sigma = 544^{+18}_{-25}~{\rm MeV}~~.
\end{equation}

Note that in the neighborhood of this singularity, the $\pi\pi$ phase shift is still small, far from the value $\pi/2$ that would constitute a lowest resonance.  The pattern of singularities in $\pi\pi$ $s$- and $p$-waves  in Fig.\ref{fig:6} shows the $\rho$ and $f_0$ resonances together with the ``$\sigma$". Obviously, this broad $\sigma$ structure does not suggest itself as the ``elementary" boson sometimes used to parametrize two-pion exchange in the nucleon-nucleon interaction.  

\begin{figure}[htb]
\centerline {
\includegraphics[width=6cm]{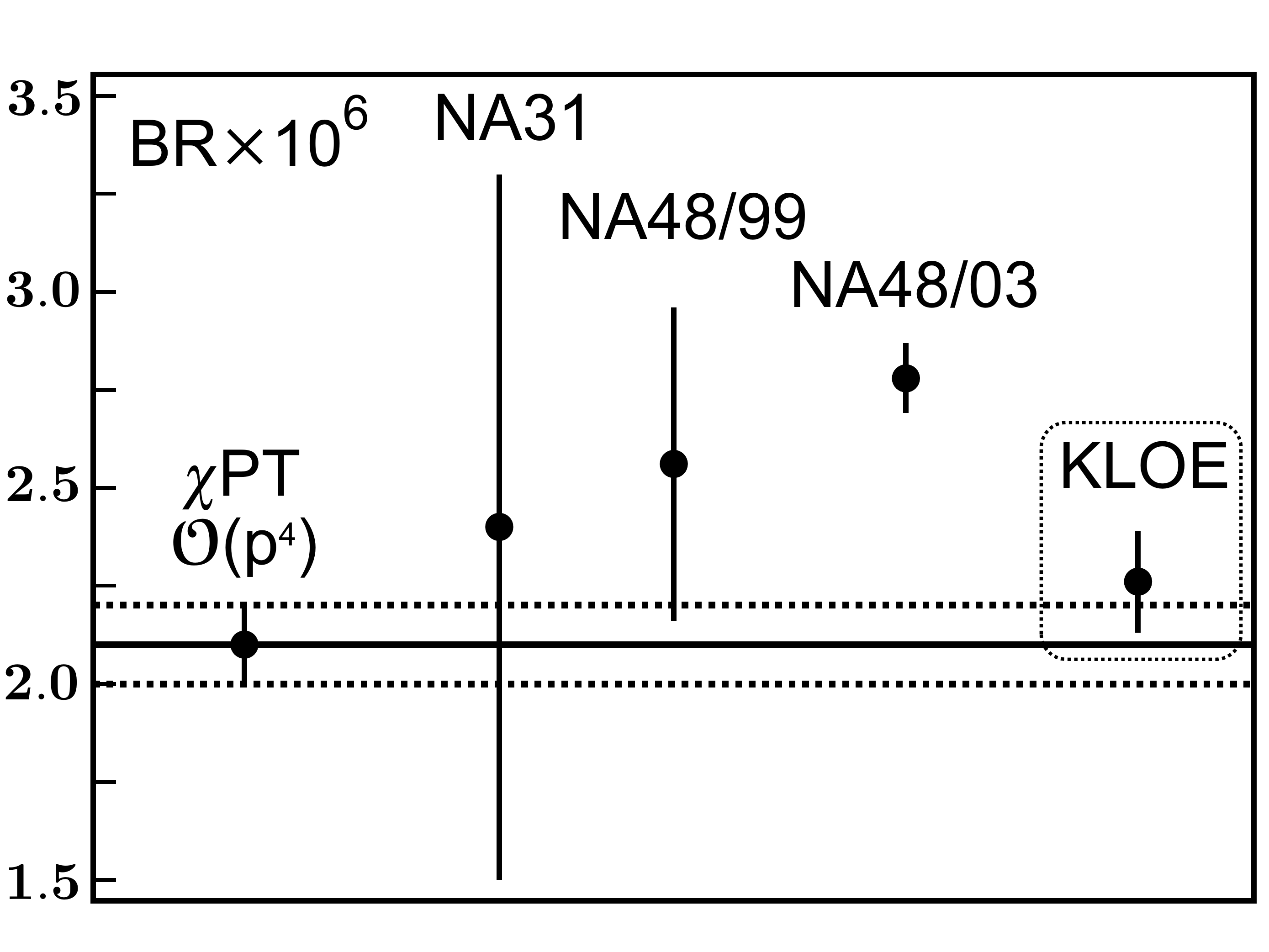}}
\caption{Branching ratio $BR(K_S\rightarrow\gamma\gamma)$ from chiral $SU(3)\times SU(3)$ perturbation theory (left) and KLOE measurement (right) \cite{kloe08}. Also shown are earlier data from NA31 and NA48. } 
\label{fig:7}
\end{figure}

The question has frequently been raised whether the mass of the strange quark, $m_s \sim 0.1$ GeV, 
is sufficiently small so as to justify a perturbative expansion within chiral $SU(3)\times SU(3)$ (see Refs. \cite{ecker08} for recent state-of-the-art assessments). One of the test cases is the $K_S\rightarrow
\gamma\gamma$ decay. There has been a long standing discrepancy between earlier measurements
from NA48 and the ${\cal O}(p^4)$ chiral perturbation theory ($\chi$PT) prediction of the branching ratio for this decay. With new data from KLOE \cite{kloe08} this discrepancy appears to be resolved. The KLOE result $BR(K_S\rightarrow\gamma\gamma) = (2.26\pm 0.12\pm 0.06)\times 10^{-6}$ is now perfectly consistent with the $\chi$PT calculation (see Fig.\ref{fig:7}).

{\it 2.2 The baryon sector.} Low-energy QCD with baryons figures as a chiral effective field theory with light and ``fast" Nambu-Goldstone bosons (the lightest pseudoscalar octet) coupled to heavy and ``slow" baryons (the spin-1/2 octet and the spin-3/2 decuplet). Given the much increased number of low-energy constants that appear in this approach, one cannot expect its predictive power and precision to match the one reached in the meson sector. Nonetheless there has been significant progress in recent years, again in conjunction with lattice QCD which now approaches quark masses as  low as 20 MeV
(or pion masses $m_\pi$ around twice the physical value). At this point chiral extrapolations begin to be reliable.

\begin{figure}[htb]
\begin{minipage}[t]{6.5cm}
\includegraphics[width=6.5cm]{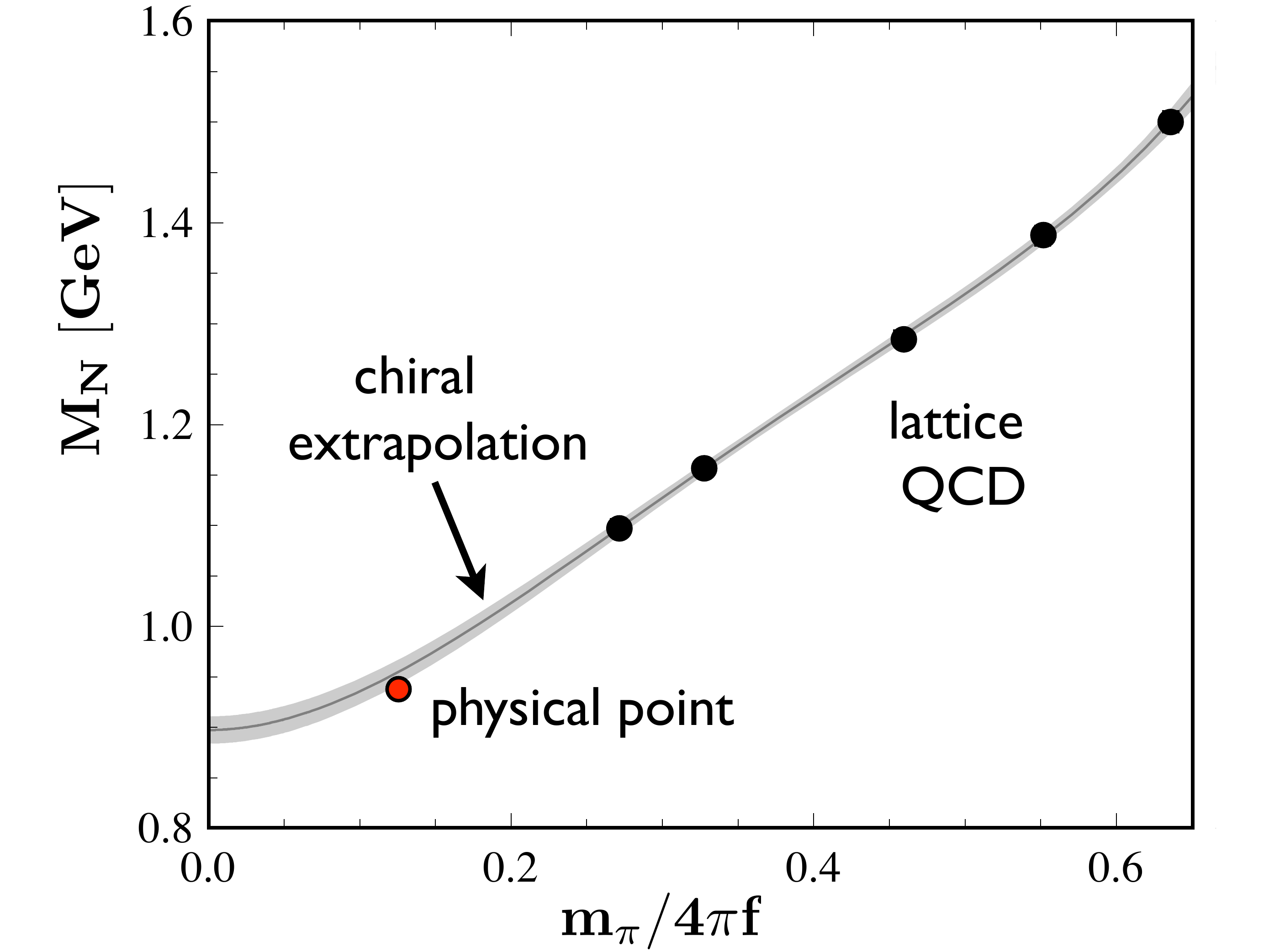}
\caption{Nucleon mass as function of pion mass from lattice QCD (full dots) and NNLO chiral SU(2) perturbation theory extrapolation via the physical point to the chiral limit ($m_\pi^2 \propto m_q \rightarrow 0$) \cite{walker-loud08}.}
\label{fig:8}
\end{minipage}
\hspace{\fill}
\begin{minipage}[t]{6.5cm}
\includegraphics[width=6.7cm]{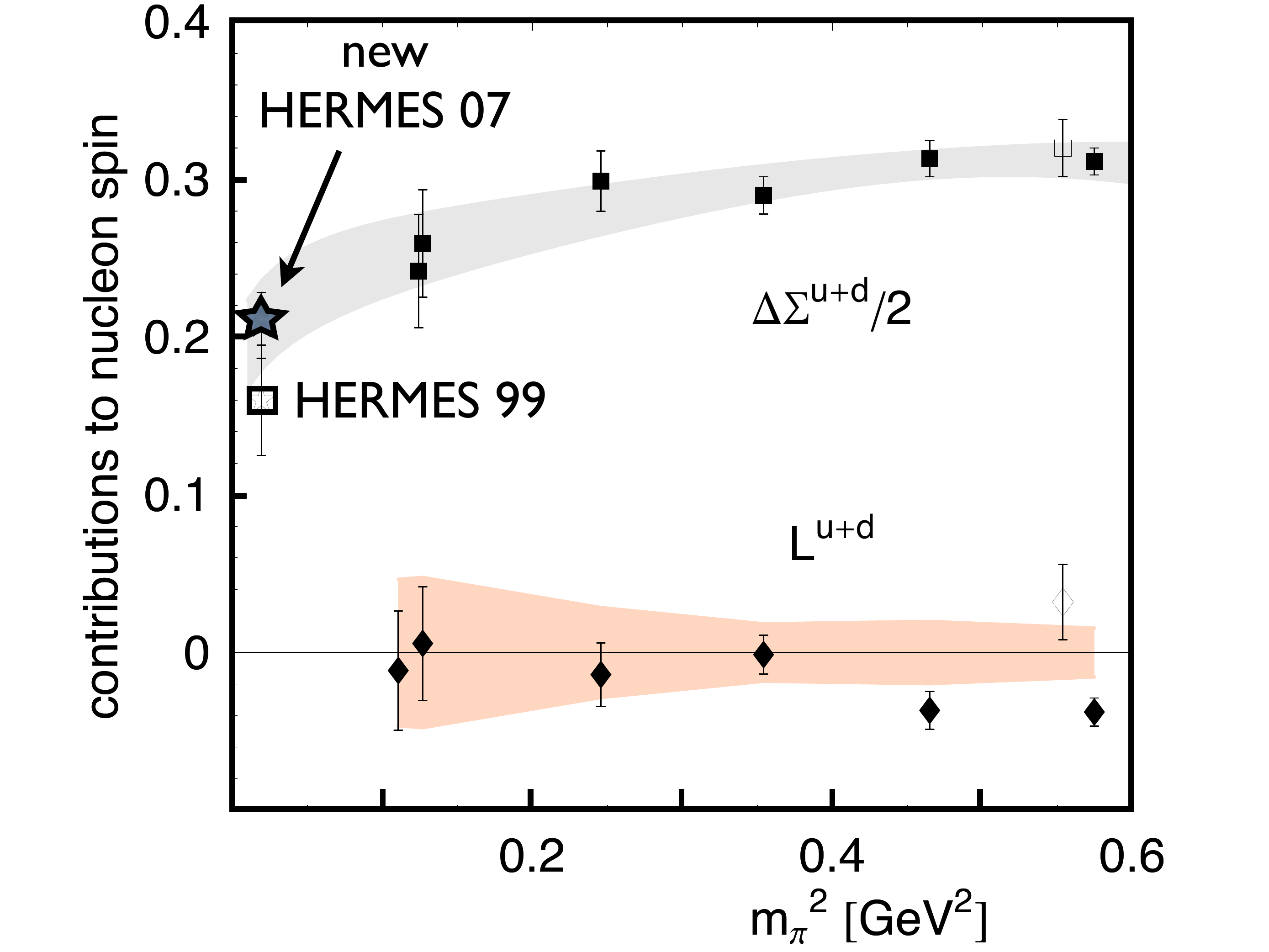}
\caption{Contributions of quark spins and orbital angular momenta to the proton spin from lattice QCD and in comparison with HERMES data (from \cite{haegler08}).}
\label{fig:9}
\end{minipage}
\end{figure}

Two representative examples (out of many) are selected here to give an impression of the progress achieved. Consider first the nucleon mass,
\begin{equation}
M_N = M_0 + \Delta M_N(m_\pi)\propto\langle N|{\beta(g)\over 2g}\,Tr(G_{\mu\nu}G^{\mu\nu}) + \sum_i m_i\,\bar{q}_i q_i|N\rangle~~,
\label{mass}
\end{equation}  
written here schematically in two complementary ways. The first part on the r.h.s. stands generically for the $\chi PT$ expansion, with $M_0$ the
nucleon mass in the chiral limit and $\Delta M_N(m_\pi)$ representing a series in powers and logarithms of $m_\pi$. The second part relates $M_N$ to the QCD trace anomaly, emphasizing the gluon-dynamical origin of the nucleon mass, plus corrections from current quark masses. Fig.\ref{fig:8} shows lattice QCD data \cite{walker-loud08} together with a chiral extrapolation based on Ref.\cite{procura06}. Such extrapolations are now performed routinely and extended to the masses of the baryon octet using chiral $SU(3)\times SU(3)$. Most recent lattice QCD results \cite{aoki08,duerr08}, with $N_f = 2+1$ quark flavours of almost physical masses, and their analysis \cite{young09} have set new standards of quantitative agreement between QCD and empirical baryon masses.

A second example is the spin structure of the nucleon. Generalized parton distributions (GPD) permit
to separate the quark spin and orbital angular momentum contributions, $J_q = \Delta\Sigma_q/2 + L_q$, to the total spin of the nucleon, $\sum_q J_q + J_{gluons} = 1/2$. Lattice QCD computations of moments of the relevant  GPD's have now been preformed \cite{haegler08}, with the results $\Delta\Sigma_{u+d}$ and $L_{u+d}$ for the proton shown in Fig.\ref{fig:9}. It is remarkable that the total $up$- plus $down$-quark orbital angular momentum appears to be compatible with zero. The flavour decomposition \cite{haegler08} gives $L_d \simeq -L_u \sim 0.2$ and $J_d = \Delta\Sigma_d/2 + L_d \simeq 0$. So one would reach the surprising conclusion that  only $u$-quarks contribute to the proton spin, qualitatively at variance with basically all models of the nucleon. The interpretation of the lattice QCD result requires, however, an evolution from the characteristic lattice scale determined by the inverse lattice spacing, $Q\sim 1/a$, down to lower $Q$. This appears to resolve the puzzle as pointed out in \cite{thomas08}.

{\it 2.3 The interface of QCD and nuclear physics.} Effective field theory combined with renormalization group methods has become a solid basis for dealing with nuclear few- and many-body problems \cite{schwenk09}.
The separation of scales characteristic of chiral effective field theory defines a systematic hierarchy of contributions to the nucleon-nucleon interaction and nuclear three-body forces \cite{epelbaum06}, driven by pions as Nambu-Goldstone bosons.  The emerging series is organised in powers of the small quantity $Q/4\pi f_\pi$, where $Q$ stands generically for low energy or momentum  and $4\pi f_\pi$ is again the spontaneous chiral symmetry breaking scale of order 1 GeV. The familiar one-pion exchange interaction comes at leading order. At next-to-leading order, ${\cal O}(Q^2)$, a first set of two-pion exchange mechanisms enters together with contact terms encoding unresolved short distance dynamics (which one can hope to ``resolve" further as lattice QCD progresses \cite{ishii07}). At the next higher order (${\cal O}(Q^3)$), more two-pion exchange processes are turned on, in particular those involving the strong spin-isospin polarizability of the nucleon as it is manifest in the $N\rightarrow \Delta(1232)$ transition that dominates $p$-wave pion-nucleon scattering. At that same order three-body interactions have their entry in a well defined book-keeping scheme.

In nuclear matter an additional relevant scale is the Fermi momentum $p_F$. Around the empirical saturation point with $p_F \simeq 0.26$ GeV, the nucleon Fermi momentum  and the pion mass, together with the $N-\Delta$ mass difference, are all comparable {\it small} scales: we have $p_F \sim 2\,m_\pi \sim M_\Delta - M_N \ll 4\pi\,f_\pi \sim 1$ GeV. This implies that at the densities of interest in nuclear physics, $\rho \lesssim\rho_0 = 2p_F^3/3\pi^2 \simeq 0.16$ fm$^{-3} \simeq 0.45\,m_\pi^3$, pions $must$ be treated as $explicit$ degrees of freedom in any meaningful description of the nuclear many-body problem. The strong pion-exchange tensor force and, in particular, two-pion exchange processes involving intermediate spin-isospin  ($N\rightarrow \Delta$) excitations, play a leading role at the distance scales characteristic of the nuclear bulk. 

In-medium chiral perturbation theory has emerged as a successful framework for low-energy pion-nucleon dynamics in the presence of a filled Fermi sea of nucleons. One- and two-pion exchange processes, treated explicitly, govern the long-range interactions at distance scales $d \gtrsim 1/p_F$ relevant to the nuclear many-body problem. Short-range mechanisms, with spectral functions involving masses far beyond those of two pions, are not resolved in detail at nuclear Fermi momentum scales and can be subsumed in contact interactions and derivatives thereof. This {\it separation of scales} argument makes strategies of chiral effective field theory work also for nuclear problems, with the {\it small} scales ($p_F, m_\pi, M_\Delta - M_N $) distinct from the {\it large} ones ($4\pi f_\pi, M_N$). Applications of this scheme have been quite successful in recent years (see e.g. \cite{weise07} for recent surveys).

In this presentation we choose to focus on a quantity at the basics of low-energy QCD and its interface with nuclear physics: the chiral (quark) condensate and its density dependence in nuclear matter. Using the Feynman-Hellmann theorem, the following expression for the ratio of
in-medium to vacuum chiral condensates (at zero temperature, $T = 0$) can be derived:
\begin{equation}
{\langle\bar{q}q\rangle_\rho\over\langle\bar{q}q\rangle_0} = 1-{\rho\over f_\pi^2}\left[{\sigma_N\over m_\pi^2}\left(1 - {3\,p_F^2\over 10\,M_N^2} + \dots\right)+ {\partial\over\partial m_\pi^2}\left({E_{int}(p_F)\over A}\right)\right]~~,
\label{cond}
\end{equation}
as a function of the baryon density $\rho = 2p_F^3/(3\pi^2)$. In this expression the first term in square brackets represents the free nucleon Fermi gas with the sigma term $\sigma_N = m_q(\partial M_N/\partial m_q) \simeq 0.05$ GeV. This term gives the leading linear density dependence plus a small correction from the kinetic energy of the nucleons. 
\begin{figure}[htb]
\centerline {
\includegraphics[width=6.5cm]{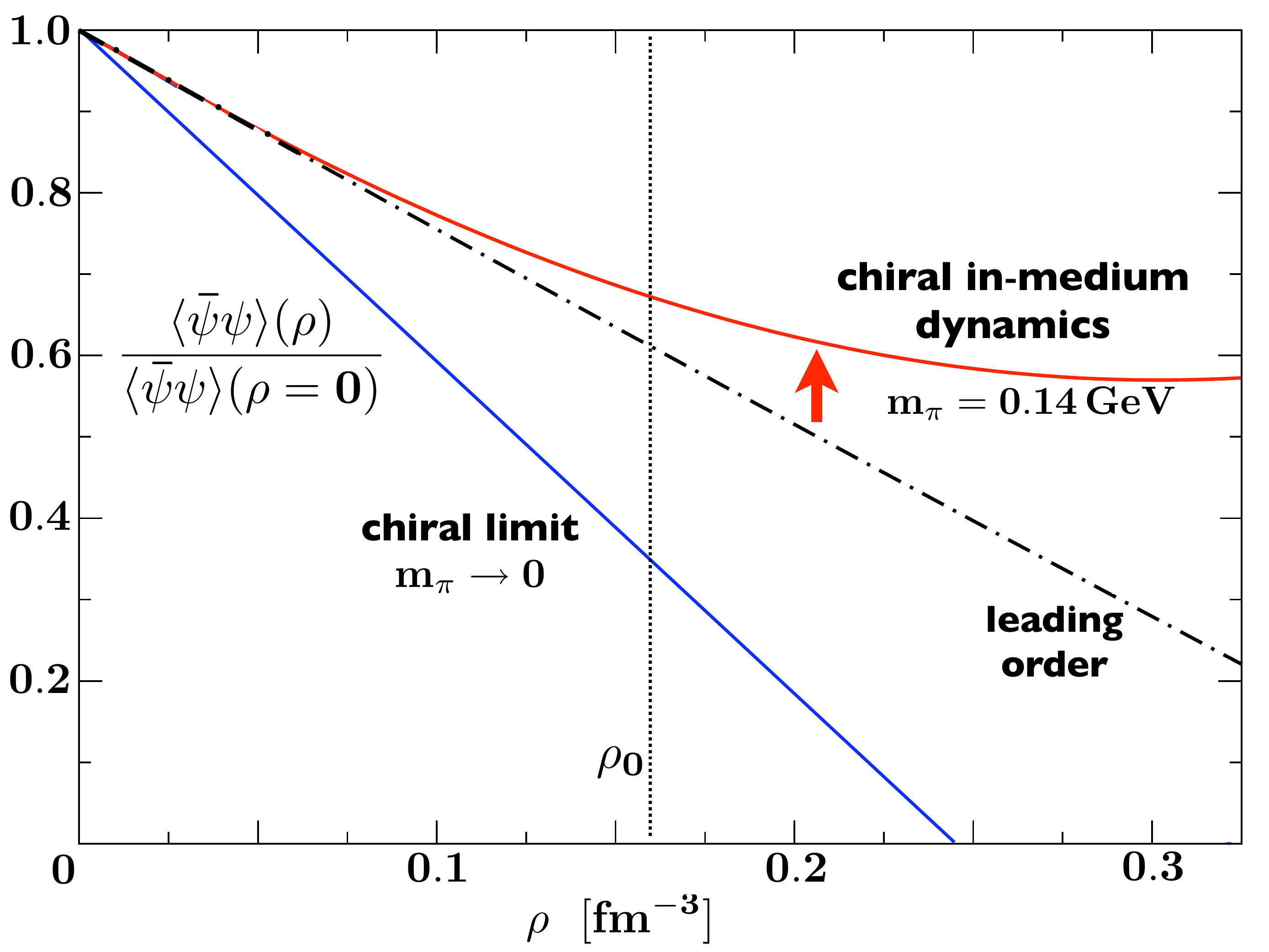}}
\caption{Density dependence of the chiral condensate \cite{kaiser08}. Dash-dottted curve: leading order term linear in $\rho$ (see Eq.(\ref{cond})) with $\sigma_N = 50$ MeV. Upper curve: result of in-medium chiral perturbation theory \cite{fritsch05} to three loop order in the energy density, using the physical pion mass. Lower curve: same calculation performed in the chiral limit ($m_\pi \rightarrow 0$).} 
\label{fig:10}
\end{figure}
Non-trivial density dependence comes from the remaining part involving the interaction energy per nucleon and its derivative with respect to the pion mass. This term can be calculated using in-medium chiral perturbation theory. In the present case the calculation is taken to three-loop order in the energy density \cite{fritsch05}, including chiral two-pion exchange dynamics and related three-body forces in the presence of the filled Fermi sea of nucleons.  

The result \cite{kaiser08} is shown in Fig.\ref{fig:10}. Up to the density of normal nuclear matter, $\rho_0 = 0.16$ fm$^{-3}$, the leading linear density dependence of $\langle\bar{q}q\rangle_\rho$ dominates, reducing the magnitude of the condensate at $\rho_0$ to about two thirds of its vacuum value. This behaviour is interpreted as the source of the strong Lorentz scalar Hartree potential which figures in relativistic nuclear mean field phenomenology. The in-medium reduction of the chiral condensate finds its correspondence in an enhancement of the $s$-wave pion-nuclear interaction near threshold which is observed in low-energy pion-nucleus scattering \cite{friedman05} and deeply bound pionic atoms \cite{suzuki04,kolomeitsev03} (see Ref. \cite{friedman07} for a recent review).

The combination of Pauli principle effects acting on chiral two-pion exchange and three-body forces is primarily responsible for the non-linear $\rho$ dependence of $\langle\bar{q}q\rangle_\rho$ observed at higher densities. The tendency towards chiral restoration is evidently delayed and not expected to occur
below twice or three times $\rho_0$. This statement turns out to depend crucially on the pion mass. In the chiral limit, $m_\pi \rightarrow 0$, the intermediate range attraction in the NN interaction is enhanced
and the transition to chiral symmetry in the Wigner-Weyl realization with vanishing quark condensate would appear at only 1.5 times $\rho_0$. Obviously, nuclear physics would be very different in the chiral limit. It is quite remarkable how much it relies on the fine tuning of explicit chiral symmetry breaking by the small, non-zero $u$- and $d$-quark masses. 

\section{Low-energy QCD with strangeness}

Chiral perturbation theory fails completely for low-energy $\bar{K}N$ interactions.
The reasons are threefold: first, the $I=0$ s-wave $\bar{K}N$ interaction derived from the chiral
$SU(3)\times SU(3)$ meson-baryon effective Lagrangian is attractive close
to the $\bar{K}N$ threshold and sufficiently strong to produce a $\bar{K}N$ bound state just
about 10 MeV below threshold. Secondly, the $\pi\Sigma$ interaction derived from the
same effective Lagrangian is also attractive and sufficiently strong to produce a broad resonance
above the $\pi\Sigma$ threshold. And thirdly, there is a strong coupling between these $\bar{K}N\leftrightarrow\pi\Sigma$ channels. This coupled-channel dynamics generates the $\Lambda(1405)$ 
at about 27 MeV below $\bar{K}N$ threshold, with a decay width of 50 MeV into $\pi\Sigma$. The non-perturbative chiral SU(3) coupled channels approach \cite{kaiser95} is a useful theoretical framework that has been applied successfully in this context. 

\begin{figure}[htb]
\begin{minipage}[t]{6.5cm}
\includegraphics[width=6.5cm]{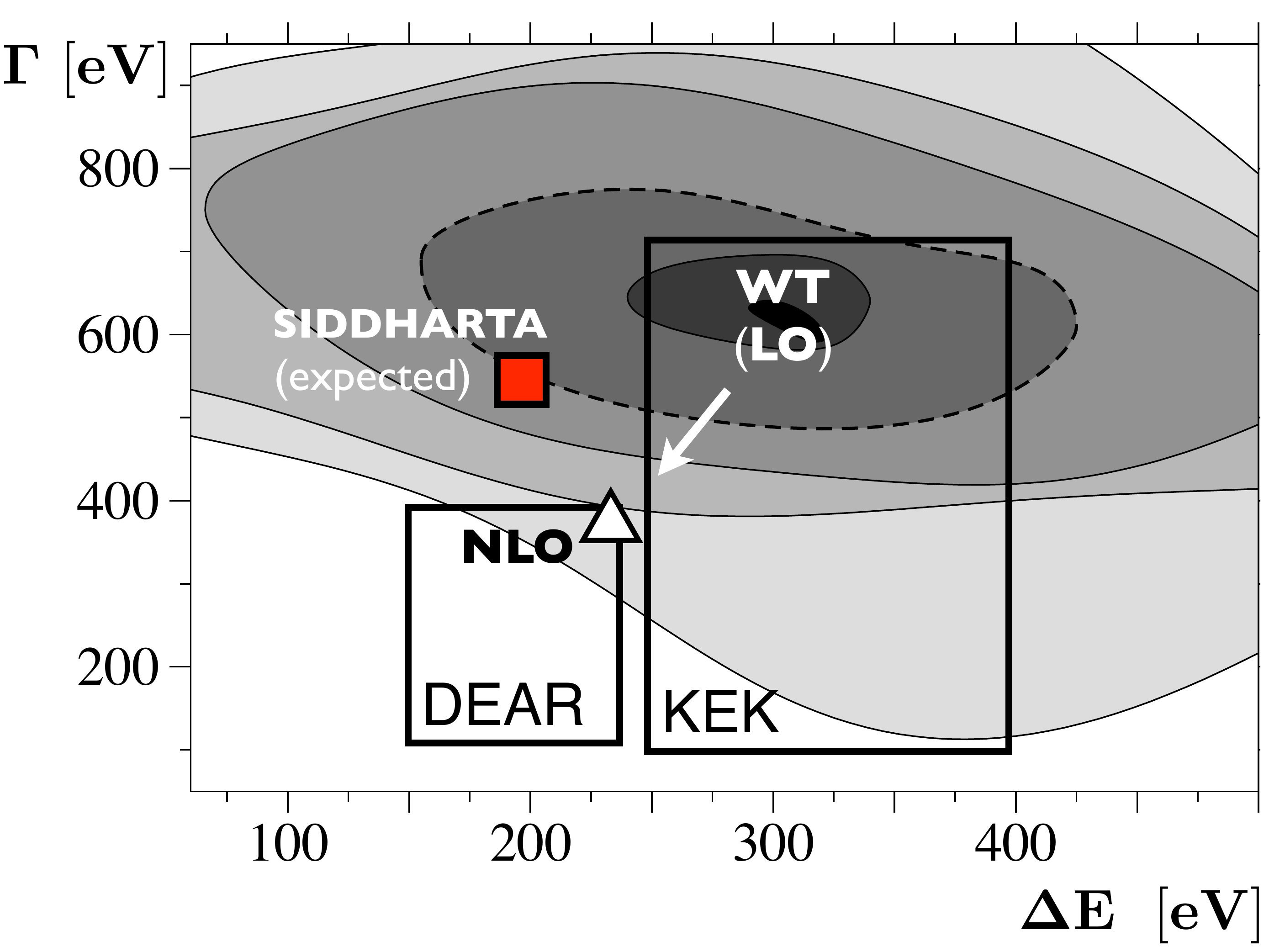}
\caption{Energy shift $\Delta E$ and width $\Gamma$ of kaonic hydrogen as determined by
the KEK \cite{kek97} and DEAR \cite{dear05} experiments. Results from chiral SU(3) coupled-channels
calculations \cite{borasoy05,borasoy06} are also shown (see text). The expected precision from the forthcoming SIDDHARTA experiment is sketched for orientation.}
\label{fig:11}
\end{minipage}
\hspace{\fill}
\begin{minipage}[t]{6.5cm}
\includegraphics[width=6.5cm]{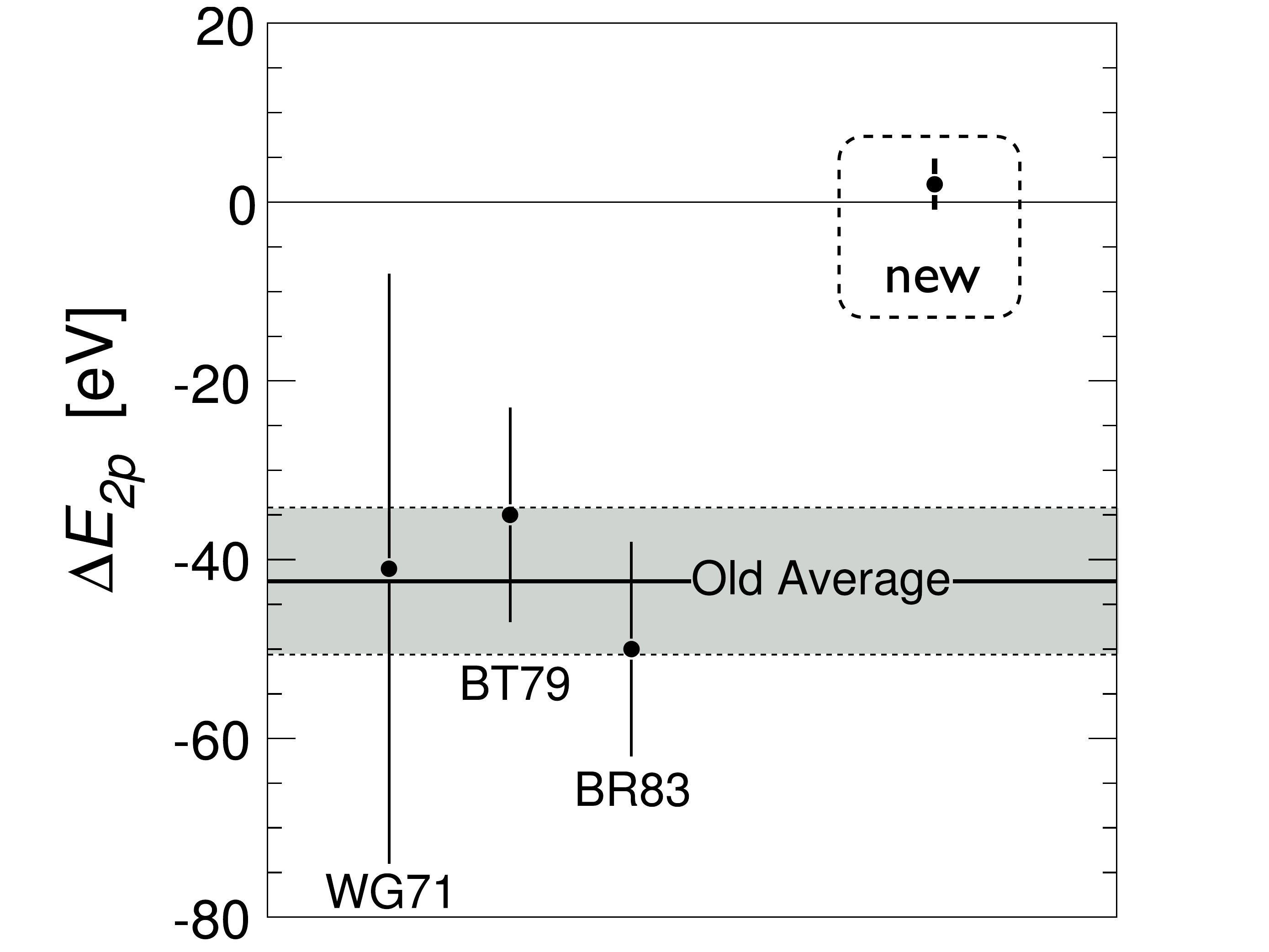}
\caption{Energy shift of the $2p$ level deduced from the $3d\rightarrow 2p$ transition in kaonic $^4$He. Previous data (``old average") are compared with the new result from a high-precision measurement at KEK \cite{okada07}.}
\label{fig:12}
\end{minipage}
\end{figure}

A basic empirical constraint for the theory is the complex $K^-p$ scattering length deduced from precision measurements of kaonic hydrogen. Experiments performed at KEK \cite{kek97} and LNF (DEAR) \cite{dear05} extracted the strong interaction energy shift and width as displayed in Fig.\ref{fig:11}. Also shown in this figure are calculations based on the 
chiral SU(3) coupled channels scheme using the leading order (LO) Weinberg-Tomozawa (WT) terms and next-to-leading order (NLO) corrections in the meson-baryon effective Lagrangian \cite{borasoy05},
together with a detailed uncertainty analysis (shaded areas
arranged from black to light grey with increasing $\chi^2$/d.o.f.) \cite{borasoy06}. These calculations reproduce the earlier KEK results together with low-energy scattering data. The more recent DEAR data can also be reproduced but at the expense of an inconsistency with the empirical $K^-p$ elastic scattering cross section.  Further clarification is required and likely to be drawn from the forthcoming SIDDHARTA experiment at LNF (Frascati) \cite{curceanu08} where one hopes to
achieve an improvement on the precision of the shift and width by about an order of magnitude. 

A recent experimental highlight is the accurate determination of the energy shift deduced from the $3d\rightarrow 2p$ transition in the kaonic $^4$He atom \cite{okada07} (see Fig.\ref{fig:12}). This resolves a long standing puzzle and places kaonic $^4$He consistently back into the systematic pattern of energy shifts observed in kaonic atoms \cite{friedman07}. The small $\Delta E_{2p}$ for kaonic  $^4$He basically rules out a ``superstrong" antikaon-nuclear interaction as it has been occasionally
advertised.

The $\bar{K}N\leftrightarrow\pi\Sigma$ coupled-channnels dynamics generated by the chiral SU(3) meson-baryon effective Lagrangian implies that the low-energy $\bar{K}N$ interaction cannot simply be described in terms of a local potential as suggested in \cite{akaishi02}. In fact, after ``integrating out" the strongly coupled $\pi\Sigma$ channel, the effective s-wave interaction in the $\bar{K}N$ channel
is complex, non-local and energy dependent \cite{hyodo08}. This has consequences for the discussion of a current hot topic, namely the quest for deeply bound, narrow antikaon-nuclear clusters \cite{akaishi02}. It requires a detailed off-shell extrapolation of the $\bar{K}N$ interaction into the far subthreshold region. Not surprisingly, such an extrapolation is necessarily limited in its predictive power. 

As a prototype for $\bar{K}$-nuclear clusters, the $K^-pp$ system has been investigated in some detail. Three-body Faddeev calculations \cite{shevchenko07,ikeda07} and variational approaches \cite{yamazaki07,dote08} suggest  $K^-pp$ bindung energies in the range $ B \sim 20 - 70$ MeV together with large widths, $\Gamma \sim 40 - 110$ MeV.
The experimental situation is so far not conclusive \cite{iwasaki08,filippi08} but the search
continues.

\section{Low-energy QCD with charm quarks}

\begin{figure}[htb]
\begin{minipage}[t]{6.5cm}
\includegraphics[width=6.7cm]{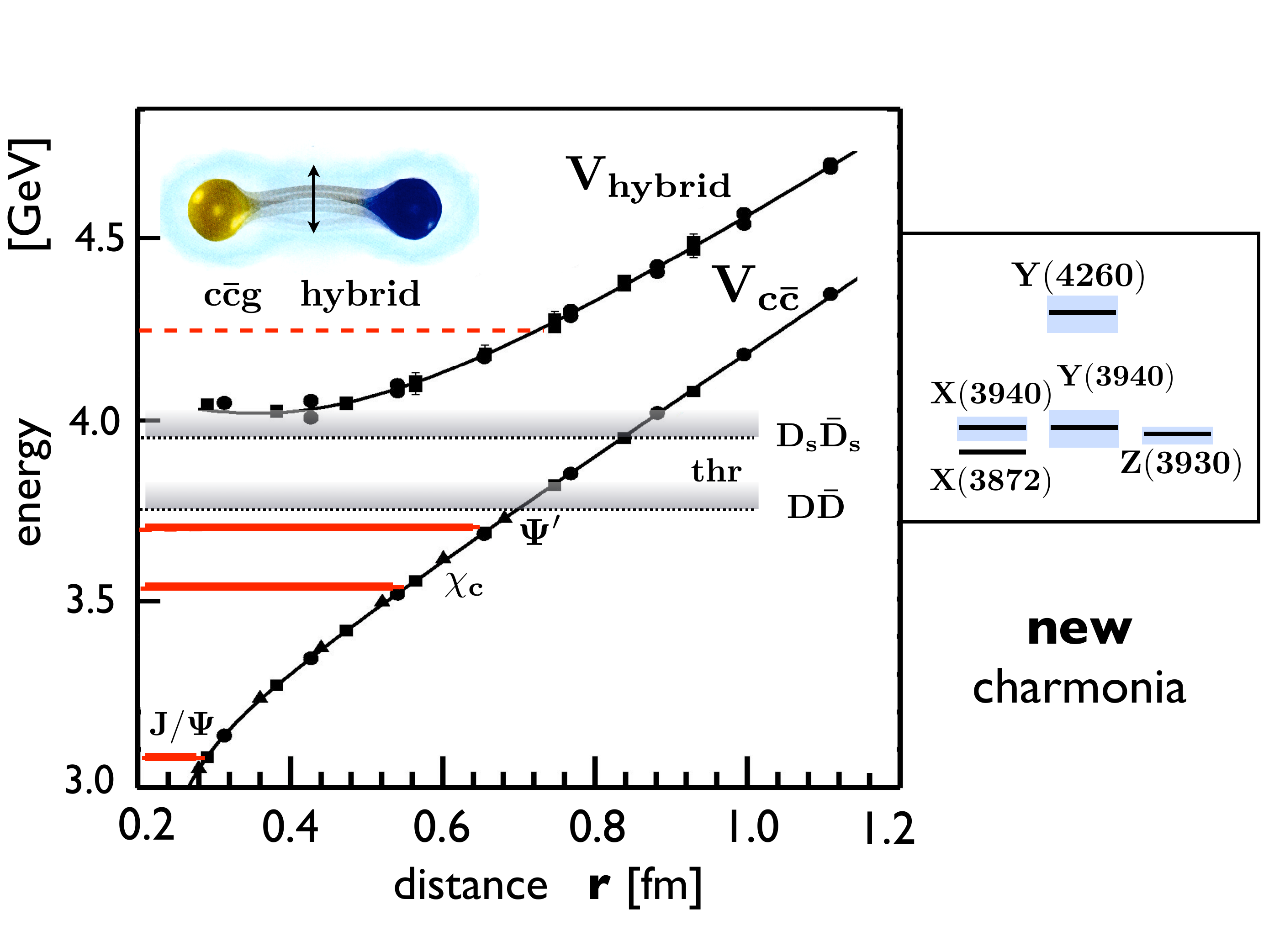}
\caption{Schematic picture of charmonia and hybrids expected from potentials
generated in Lattice QCD \cite{bali00}. Open charm thresholds are indicated together with some of the ``new" charmonium states.}
\label{fig:13}
\end{minipage}
\hspace{\fill}
\begin{minipage}[t]{6.5cm}
\includegraphics[width=6.7cm]{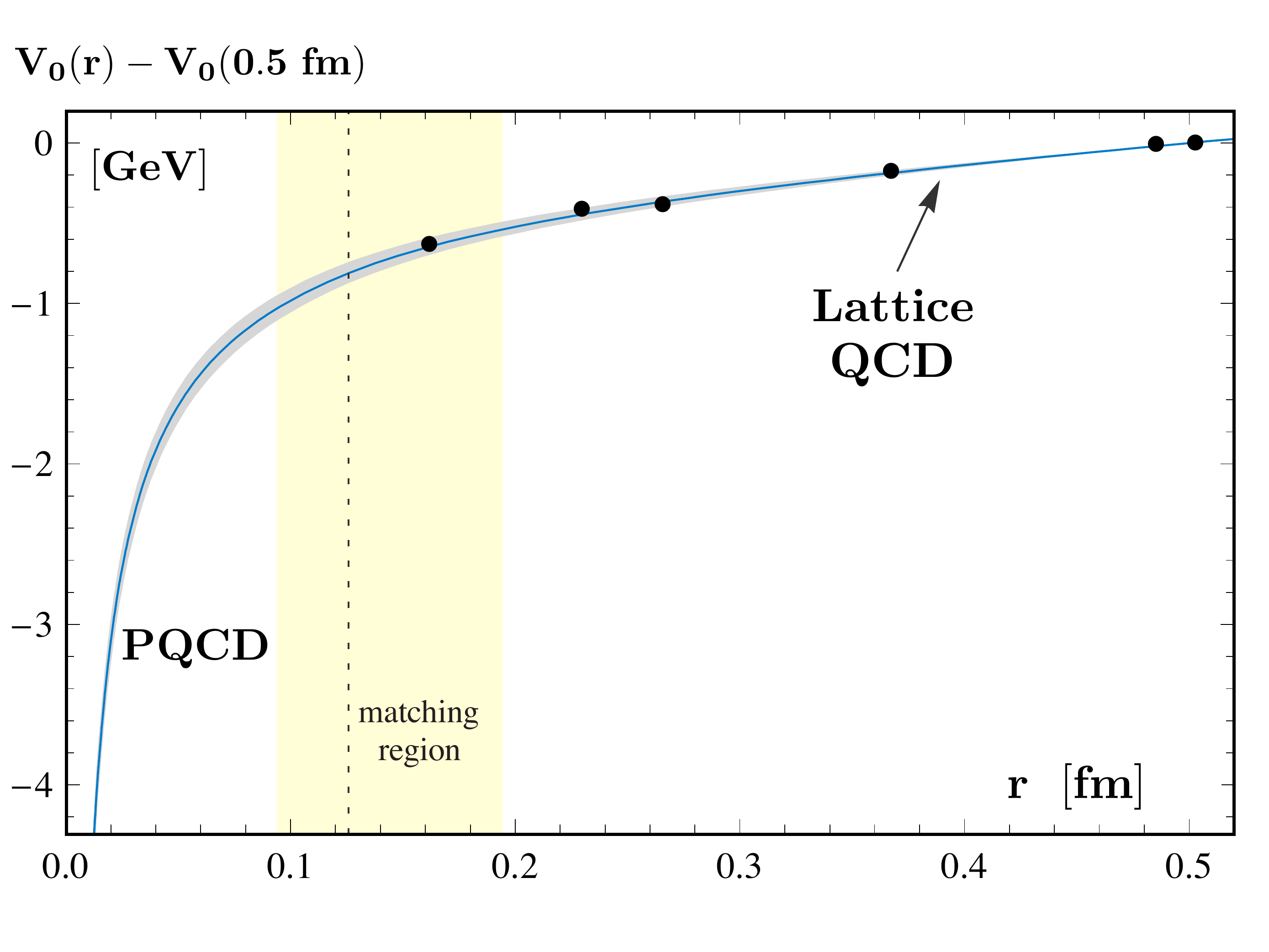}
\caption{Matching of the static quark-antiquark potential between perturbative QCD at short distance and lattice QCD \cite{koma07} at larger distance (from \cite{laschka09}).}
\label{fig:14}
\end{minipage}
\end{figure}

Charmonium physics is presently experiencing a remarkable renaissance. Since 2002 a dozen of new states have been found by Belle, BaBar, CLEO-c, CDF and D0. All of these states are embedded in
the open-charm continuum (see Fig.\ref{fig:13}). Unravelling their detailed structure in terms of their
Fock space decompositions, generically written as $|X\rangle = a_1|c\bar{c}\rangle + a_2|[c\bar{q}][\bar{c}q]\rangle + a_3|[cq][\bar{c}\bar{q}]\rangle + a_4|c\bar{c}\,glue\rangle + \dots$ ,  is one of the outstanding challenges in hadron physics. This will require efforts of combining coupled-channels methods \cite{barnes08} with frameworks such as heavy-quark effective field theories and non-relativistic QCD \cite{brambilla05}.

At the same time lattice QCD is making steady progress in establishing accurate results for the quark-antiquark potential beyond the static limit, i.e. in an expansion in inverse powers of the heavy quark mass \cite{koma07}. Matching these potentials with perturbative QCD at short distance is a subject that has already a history but can now be systematically updated. An example of such an accurate
matching is shown for the leading (static) potential in Fig.\ref{fig:14}. 

Of course the standard charmonium potential, even with inclusion of $1/M^2$ corrections, does not properly describe the physics around and above open-charm thresholds. The confinement part, linearly rising with distance $r$, is disrupted by coupled-channel dynamics which generates a complex effective $c\bar{c}$ potential. Its imaginary part represents the coupling to $D\bar{D}, D_s\bar{D_s},
D^*\bar{D}^*$ etc., while the associated dispersive real parts produce mass shifts which are by no means negligible \cite{barnes08}.  This is one out of many challenges revived by the charmonium renaissance.

\section{Summary}
\begin{itemize}
\item{Low-energy QCD with light ($u$- and $d$-) quarks in its meson (Nambu-Goldstone boson)
sector is established as a quantitative science, through the joint efforts of chiral effective field theory, lattice QCD and precision experiments. Chiral perturbation theory including strange quarks is likewise successful but with slower convergence.}
\item{Chiral effective field theory with baryons can now be considered an appropriate framework for dealing with low-energy nucleon structure and nuclei.}
\item{New and challenging phenomena in antikaon-nucleon and -nuclear systems are under investigation.}
\item{The unexpected charmonium renaissance has opened a new field of hadronic research with a high discovery potential.}
\end{itemize} 
%
%
%

%
\end{document}